\documentclass[journal]{IEEEtran}

\usepackage[cmex10]{amsmath}
\usepackage{siunitx}
\usepackage[font=footnotesize]{caption}
\usepackage{psfrag}
\usepackage[utf8]{inputenc}
\usepackage[T1]{fontenc}
\usepackage{amsmath,amsfonts,amsbsy,amssymb}
\usepackage{mathabx}
\usepackage{mathrsfs}
\usepackage[nolist]{acronym}
\usepackage{tabularx}
\usepackage{amssymb}
\usepackage{pdfpages}
\usepackage{amsmath}
\usepackage{graphicx}
\usepackage{soul}
\usepackage{cite}
\usepackage{wasysym}
\usepackage{float}
\usepackage{xcolor}
\usepackage{subfigure}
\usepackage{ragged2e}
\usepackage{makecell, multirow, tabularx}
\usepackage{booktabs}

\usepackage{siunitx, mhchem}
\usepackage{xcolor}
 
\usepackage[ruled, lined, longend, linesnumbered]{algorithm2e}

\usepackage{xcolor}
\usepackage{url}
\usepackage{multicol}
\usepackage{verbatim}
\usepackage{array}
\definecolor{uranianblue}{rgb}{0.69,0.86,0.96}
\definecolor{thistle}{rgb}{0.85,0.75,0.85}
\definecolor{mintgreen}{rgb}{0.6,1.0,0.6}
\definecolor{new}{rgb}{0.4940 0.1840 0.5560}
\definecolor{azurexwebcolor}{rgb}{0.94,1.0,1.0}

\newcommand\cparagraph[1]{\vspace{0.6mm}\noindent\textbf{#1.}}

\ifCLASSINFOpdf
\else
\fi
\begin{document}
%
\title{Privacy-Aware Smart Cameras: View Coverage via Socially Responsible Coordination}
%
%
%

\author{\IEEEauthorblockN{Chuhao Qin\IEEEauthorrefmark{1}, Lukas Esterle\IEEEauthorrefmark{2} and 
Evangelos Pournaras\IEEEauthorrefmark{1}} \\
\IEEEauthorblockA{
\IEEEauthorrefmark{1}School of Computer Science, University of Leeds, UK\\
\IEEEauthorrefmark{2}Department of Electrical and Computer Engineering, Aarhus University, Denmark
}
}


\maketitle

\begin{abstract}
Coordination of view coverage via privacy-aware smart cameras is key to a more socially responsible urban intelligence. Rather than maximizing view coverage at any cost or over relying on expensive cryptographic techniques, we address how cameras can coordinate to legitimately monitor public spaces while excluding privacy-sensitive regions by design. This article proposes a decentralized framework in which interactive smart cameras coordinate to autonomously select their orientation via collective learning, while eliminating privacy violations via soft and hard constraint satisfaction. The approach scales to hundreds up to thousands of cameras without any centralized control. Experimental evidence shows $18.42\%$ higher coverage efficiency and $85.53\%$ lower privacy violation than baselines and other state-of-the-art approaches. This significant advance further unravels practical guidelines for operators and policymakers: how the field of view, spatial placement, and budget of cameras operating by ethically-aligned artificial intelligence jointly influence coverage efficiency and privacy protection in large-scale and sensitive urban environments.
\end{abstract}


%
\IEEEpeerreviewmaketitle

\section{Introduction}
Smart cameras are now deeply embedded in modern cities. They monitor traffic congestion, enhance public safety, and can even recognize human activity in public spaces~\cite{natarajan2015multi}. Although cities deploy more cameras to ``cover everything'', privacy concerns cannot be ignored anymore. So far, cameras are not made to be socially responsible, since they can be misused, hacked, or repurposed~\cite{video_demo_2026} not for good. Surveillance systems, including CCTV, can be used to suppress citizens’ freedom, censor, control and maniputale people's personal life. The risks are real: A 2022 survey reported that nearly half of retail organizations in the UK had been fined for violating General Data Protection Regulation (GDPR) regulations~\cite{secure2026}. These realities raise a fundamental question: how can cities benefit from smart cameras without compromising public trust?

In this article, we address a key challenge: \textit{Given a fixed number of cameras and their locations, how can they be coordinated to monitor an area effectively while avoiding the coverage of privacy-sensitive regions by design?} We refer to this capability as the privacy-aware multi-camera coverage coordination problem, which is approached by studying two types of constraints: (1) Soft constraints which ensure cameras to avoid overlap (excessive coverage) and blind spots (insufficient coverage); (2) Hard constraints which guarantee that cameras never observe certain areas, such as residential windows, schoolyards, or other privacy-sensitive spaces. Unlike traditional privacy-preserving approaches that rely on computationally expensive and post-hoc cryptographic techniques to encrypt captured video~\cite{ma2018camera}, our approach prevents privacy intrusion at its source by coordinating where cameras observe in the first place, which is more cost-effective and practical in real implementations. Moreover, via a socially responsible coordination, we may not need at first place excessive number of cameras, which reduces further down risks of privacy violations. 

To achieve this, we introduce a novel decentralized smart camera system. Instead of streaming all video to a central control room, cameras process information locally, exchange lightweight event-driven summaries with other cameras in short proximity, and store raw data at the edge. This greatly reduces network congestion and infrastructure costs~\cite{taj2011distributed}. The decentralized system is also scalable, i.e., new cameras can be added without redesigning the entire architecture, and resilient, i.e., since there is no single point of failure that could break the whole network~\cite{mukherjee2021decentralized}. However, new challenges are introduced by decentralization. Without a central controller, cameras need to coordinate among themselves to maintain optimized coverage while respecting privacy constraints. This requires additional local computation and frequent communication with neighboring cameras to adjust orientations dynamically. The problem becomes even harder when multiple orientation options are disregarded due to privacy violations, i.e., each camera is left with only a few feasible alternatives, making coordination more complex and potentially reducing coverage quality. To handle these challenges, a powerful and scalable coordination algorithm is essential, which should also be privacy-preserving by design.

At the core of this system is a fully decentralized collective learning algorithm, I-EPOS, the Iterative Economic Planning and Optimized Selections~\cite{pournaras2018decentralized}. This algorithm is chosen for its remarkable scalability and efficiency in coordinating thousands of agents with low computational and communication cost~\cite{pournaras2018decentralized,majumdar2023discrete}. In our framework, each camera is interactive and behaves in a ``smart'' manner: it can generate multiple possible orientation plans that respect citizens' privacy via hard constraint satisfaction, communicate with neighboring cameras, and autonomously select the optimal plans via a hierarchical coordination. Extensive experimentation validates the effectiveness of the proposed approach and addresses several questions for real-world deployments: Can the system scale to a large number of cameras? How many orientation options does each camera actually need? How well can privacy be protected under different privacy-sensitive areas? And how should operators determine the field of view, number, and location of cameras to achieve an effective balance between coverage quality, privacy protection, and deployment cost?

In summary, this article includes the following contributions: (1) The first study on privacy-aware multi-camera coordination of view coverage that prevents privacy violations by design. (2) A novel decentralized coordination framework that leverages collective learning (I-EPOS), enabling hundreds of cameras for coordinated orientation selection, achieving more efficient view coverage and better privacy protection than baseline methods. (3) New quantitative insights into the cost-benefit trade-offs of smart camera placement, offering practical guidance for deploying legitimate CCTV camera systems in urban environments that balance effective coverage with anti-surveillance capability and stronger protection of citizens’ privacy. (4) An open dataset\footnote{Available at: https://doi.org/10.6084/m9.figshare.31332856} of generated camera orientation plans to support reproducibility and encourage further research in privacy-aware smart cameras.

\newcommand{\tabincell}[2]{\begin{tabular}{@{}#1@{}}#2\end{tabular}}
\begin{table*}[!t]
\centering
\caption{Comparison of approaches for discrete-choice combinatorial optimization problem.}
\centering
\footnotesize
\label{tab:compare}
    \begin{tabular}{lllll}
    \hline
    Attributes	&	I-EPOS~\cite{pournaras2018decentralized,majumdar2023discrete}	&	EPOS~\cite{pournaras2013multi} & COHDA~\cite{bremer2017agent}  & GGV~\cite{suresh2022efficient}	\\ \hline
    Computational cost &
    \tabincell{l}{agent: $O(KL)$; \\ system: $O(KL\ log\ U)$}
    &
    \tabincell{l}{agent: $O(K^C)$; \\ system: $O(K^C\ log\ U)$}
    &
    \tabincell{l}{agent: $O(KL)$; \\ system: $O(KL)$}
    &  
    \tabincell{l}{agent: $O(KL)$; \\ system: $O(KLU)$} \\ \hline
    
    Communication cost &
    \tabincell{l}{agent: $O(L)$; \\ system: $O(L\ log\ U)$}
    &
    \tabincell{l}{agent: $O(K)$; \\ system: $O(K\ log\ U)$}
    &
    \tabincell{l}{agent: $O(UL)$; \\ system: $O(UL)$}
    &  
    \tabincell{l}{agent: $O(UL)$; \\ system: $O(UL)$} \\ \hline

    Scalability & \checkmark & \checkmark & $\times$  & $\times$ \\ \hline
    
    Privacy awareness & \checkmark & $\times$ & $\times$  & $\times$ \\ \hline
    \multicolumn{5}{p{15cm}}{ $K$: number of plans (options), $L$: number of iterations, $C$: number of children, $U$: number of agents} \\
    \multicolumn{5}{p{15cm}}{Criteria covered $\checkmark$ or not $\times$}
    
\end{tabular}
\end{table*}

\section{RELATED WORK}
The key research problem with multi-camera coverage coordination addressed in this article is typically formulated as an NP-hard 0-1 discrete-choice combinatorial optimization problem. A practical example of this problem arose in intelligent transportation systems, where smart cameras are coordinated to function as distributed sensors for estimating traffic density and detecting congestion in real time~\cite{hu2023turning,guastella2023cooperative}. Deng \textit{et al.} further extended this problem to an air–ground surveillance sensor network, where aerial (drones) and ground (cameras) sensors collaborate through edge computing to coordinate coverage and enable efficient real-time target tracking~\cite{deng2021air}.   

Greedy optimization methods are simple and highly efficient for camera coverage problems~\cite{zhao2013approximate,altahir2017optimizing}. They typically assign cameras sequentially, following a raster scan order from the top-left to the bottom-right of the map, and, at each location, select the orientation that locally maximizes coverage. Esterle~\cite{Esterle2018ICASSPiot} proposed ARES and particle swarm optimization for self-organized coverage optimization in distributed camera networks. Although ARES works well, the computational and communication overhead limits the applicability of these approaches in larger camera settings. Suresh \textit{et al.}~\cite{suresh2022efficient} proposed a new Greedy Grid Voting algorithm (GGV), a two-stage grid voting algorithm for camera coverage optimization. The GGV algorithm is not constrained by raster scan order but gives more preference to covering critical regions. However, these greedy methods prioritize immediate coverage gains and rarely account for no-surveillance zones (e.g., private households, restrooms), reflecting a deployment strategy that improves coverage efficiency but neglects a more socially responsible operation.

The research of balancing coverage optimization and citizen's privacy protection has widely studied~\cite{ma2018camera}. Previous privacy-preserving camera systems, such as TargetFinder~\cite{khazbak2019targetfinder} and SecureCam~\cite{aribilola2022securecam}, used cryptographic techniques combined with video processing to protect sensitive information after images are captured. While effective in limiting direct privacy leakage, these approaches are centralized and designed for relatively small-scale deployments. They depend on a computational infrastructure to process, encrypt, and manage video streams, which is post-hoc rather than during operations or by design. This means that sensitive data collected by cameras are still accessible to third parties, violating citizens' privacy. In contrast, our framework eliminates the need for heavy cryptographic processing, ensuring scalable, resilient, and ethically aligned camera operations.

On the review of decentralized coordination approaches, collective learning solves a large class of discrete-choice combinatorial optimization problems in a decentralized manner~\cite{guastella2023cooperative,suran2020frameworks}. It is a decentralized optimization approach where autonomous agents coordinate their decision making to schedule operations or allocate resources. The two main decentralized approaches in the literature are COHDA~\cite{bremer2017agent} and EPOS~\cite{pournaras2013multi}. COHDA generalizes well in different communication structures among agents that have a full view of the system, while EPOS focuses on hierarchical acyclic graphs such as self-organized trees to perform cost-effective decision making and aggregation of choices. As COHDA shares full information between agents, it has higher communication overhead. The computational cost is lower at the global level for COHDA compared to EPOS because of the agents' brute force search to aggregate choices. As an improvement of EPOS, Pournaras \textit{et al.}~\cite{pournaras2018decentralized,majumdar2023discrete} proposed I-EPOS to significantly reduce both communication and computational cost and ensure higher cost-effectiveness than COHDA (see Table~\ref{tab:compare}). This approach aligns well with the requirements of the large-scale coverage coordination problem.

\begin{figure*}[!t]
\centerline{\includegraphics[width=\linewidth]{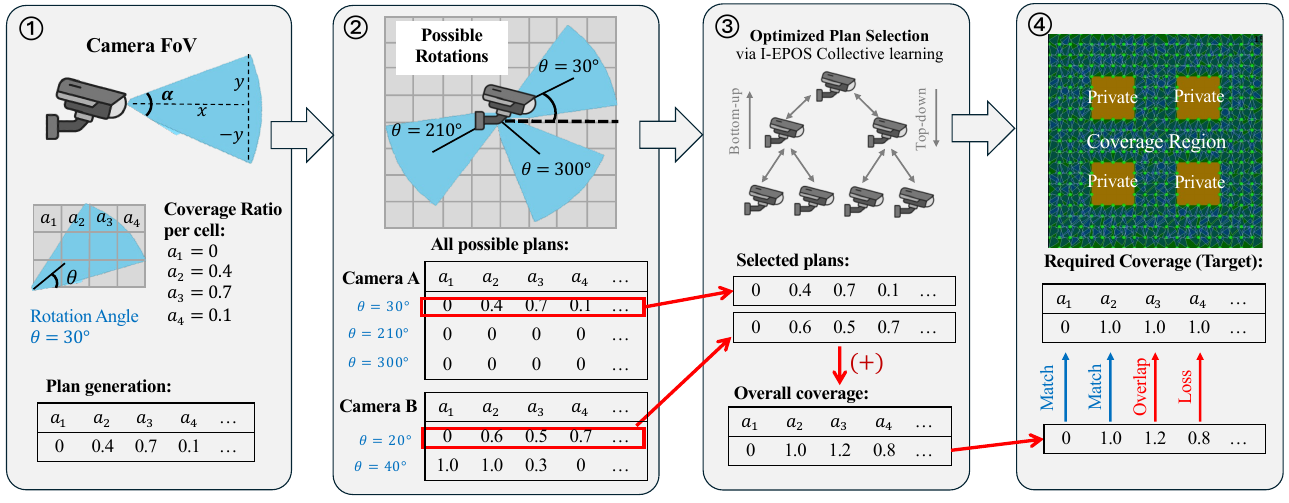}}
\caption{Overview of the smart camera coordination model. Camera A has three orientation plans ($30^{\circ}$, $210^{\circ}$, and $300^{\circ}$). When camera A rotates $30^{\circ}$, its field of view (FoV) covers $40\%$ of cell $a_2$, $70\%$ of cell $a_3$ and $10\%$ of cell $a_4$. It generates all possible $360^{\circ}$ plans and then coordinates with other cameras to observe the required regions while excluding the private regions through collective learning. The objective is to match the overall coverage to the target at each cell.}\vspace*{-5pt}
\label{fig:scenario}
\end{figure*}

\section{DECENTRALIZED COORDINATION METHOD}
This section first models the proposed problem. Then, the collective learning approach is introduced to solve the problem.

\subsection{Problem Model}
The core problem of this work is: \textit{given a set of deployed interactive smart cameras, how can they coordinate autonomously without central control to choose their viewing orientations in order to collectively maximize coverage while avoiding viewing private regions?} In this article, we model the coverage coordination problem in a 2D setting for simplicity, whereas the proposed approach can also extend to 3D environments with full pan-tilt camera control.

Consider a set of static smart cameras $\mathcal{U} \triangleq \{1,2,...,U\}$ uniformly distributed in a grid --- a 2D map (see Fig.\ref{fig:scenario}). The camera coverage is modeled deterministically using an enhanced pinhole camera model~\cite{altahir2017optimizing}. Therein, a triangular coverage area is considered with the coordinates of $(x, y)$ and $(x, -y)$. The field of view $F$ and its horizontal angle $\alpha$ can be calculated as follows:
\begin{equation}
    \tan(\frac{\alpha}{2}) = \frac{w_\text{I}}{2f} = \frac{y}{x},
    \label{eq:angle}
\end{equation}
\begin{equation}
    F = x^2 \cdot \frac{w_\text{I}}{2f},
    \label{eq:field_view}
\end{equation}
where $x$ is the effective range of a camera, $w_\text{I}$ is the width of the image sensor, and $f$ is the focal length of the camera (the unit is meter). The visibility of a point from a camera viewpoint is analyzed using ray tracing algorithms. Any segment of the ray that extends beyond the point of intersection with the boundary of an occluding object is considered invisible to the camera. This article uses Bresenham’s line algorithm for ray tracing to determine point visibility~\cite{suresh2022efficient,altahir2017optimizing}.

Each camera, denoted as $u \in \mathcal{U}$, has multiple \textit{orientation plans} $\mathcal{K}_u$ from which it can choose. Each plan $p_{uk} \in \mathcal{K}_u$ represents a possible coverage of the camera $u$ when it rotates an angle, where $k$ denotes the index of a plan and the total size of $\mathcal{K}_u$ is denoted as $K$. In addition, the whole map is divided into a number of unit areas (i.e., the cells), denoted as $\{a_1,...,a_n, ..., a_N\}$, where $n$ denotes the index of a cell, $n \leq N$. Here, the plan is defined as a vector with the elements of $p_{ukn}$ to show the coverage ratio per cell, i.e., $p_{uk} = [p_{ukn}]^N_{n=1}$, $0 \leq p_{ukn} \leq 1$. The coverage ratio at $a_n$ is calculated as follows:
\begin{equation}
    p_{ukn} = \frac{C_n(L_u, F, \theta_{uk})}{A_n},
    \label{eq:coverage_ratio}
\end{equation}
where $A_n$ is the area of cell $a_n$, and $C_u(\cdot)$ denotes the function to calculate the area of $a_n$ covered by camera $u$ based on its location $L_u$, the field of view $F$ and the rotation angle $\theta_{uk}$ corresponding to the plan of $k$.

Furthermore, the coverage requirement (i.e., target) of the map is defined. The map is divided into two types of areas: \textit{coverage regions} where each cell is expected to be observed by cameras, and \textit{private regions}, where observation is strictly prohibited. We use the target vector $T$ with the elements of $T_n = \{0,1\}$ to express the coverage requirement, where $T_n = 0$ requires not to cover the cell $a_n$ and $T_n = 1$ by one camera. Note that $T_n > 1$ is undesirable as it means excessive coverage.

To meet the coverage requirements, each camera $u$ should select an optimal plan from its coverage plans $\mathcal{K}_u$ such that the aggregated plan $G$, i.e., the overall coverage by all cameras, matches the target. $G$ is also a vector with the elements of $G_n$, which is expressed as follows:
\begin{equation}
    G_n = \sum_{u=1}^U p_{ukn} \cdot x_{uk},
\end{equation}
where $x_{uk}$ is a binary decision variable that takes $1$ if camera $u$ chooses the plan $p_{uk}$ and $0$ otherwise, $\sum_{k=1}^K x_{uk} = 1$. We define $G_n > T_n$ as \textit{overlap} and $G_n < T_n$ as \textit{loss}, $\forall n \in \mathcal{N}$.

Given the target $T$ and the aggregated plan $G$, the optimization model is formulated with both soft and hard constraints. The soft constraints aim to minimize the mismatch, such as root mean square error (RMSE), between $T$ and $G$. The hard constraints are satisfied by restricting cameras from covering any cell in private regions. To achieve the privacy-aware coverage optimization, the model is formulated as follows:
\begin{equation}
    \mathop{\min}\limits_{x_{uk}} \sqrt{\frac{\sum_{n=1}^N (G_n - T_n)^2}{N}}.
    \label{eq:inefficiency}
\end{equation}
Subject to
\begin{equation}
    G_n = T_n = 0, \; n \in \mathcal{P}.
    \label{eq:constraint_private}
\end{equation}
Here, the objective function of Eq.(\ref{eq:inefficiency}) is to minimize the coverage inefficiency. Hard constraint of Eq.(\ref{eq:constraint_private}) restricts the coverage on private regions, where $\mathcal{P}$ denotes the set of cells within private regions, $\mathcal{P} \subset \mathcal{N}$.

Note that the higher coverage inefficiency leads to both higher overlap and loss. For example, see Fig.\ref{fig:scenario}, cameras have higher overall coverage value than the target at $a_3$ (overlap), whereas $a_4$ is only $80\%$ covered (loss). This problem is shown earlier to be an NP-hard discrete-choice combinatorial optimization problem in this context~\cite{pournaras2018decentralized}. 

\begin{algorithm}[!t]
    \DontPrintSemicolon
	\caption{Plan generation strategy for each camera.} 
	\label{algorithnm1}
	\textbf{Input:} The index of camera $u$, the location $L_u$, the field of view $F$, the number of plans to generate $K$, and the map divided by $N$ cells.\;
	Initialize $\mathcal{K}_u = \emptyset$ \;
	\For{each plan index $ := 1,...,K$}
	{
        Calculate rotation angle $\theta _{uk}= k \cdot \frac{360}{K}$\;
        Find $\{C_n(L_u, F, \theta_{uk}), n \leq N\}$\;
        Calculate $p_{ukn}$ via Eq.(\ref{eq:coverage_ratio})\;
        \If{$\sum_{n=1}^N p_{ukn} > 0$} {
            \If{$\sum_{n=1}^N p_{ukn} \cdot (1 - T_n) < V $} {
                Generate the plan $p_{uk} = [p_{ukn}]^N_{n=1}$\;
                $\mathcal{K}_u = \mathcal{K}_u \cup p_{uk}$\;
            }
        }             
	}
    \textbf{Output:} The generated plans $\mathcal{K}_u$.
\end{algorithm}

\subsection{Collective Learning Approach}
The NP-hard problem is challenging to solve at scale. When coordinating a high number of cameras (e.g., one hundred), the system needs to process an enormous number of view combinations (i.e., $360^{100}$) when adjusting $1$ degree each time, which requires significant computational and communication resources. Decentralized heuristics can efficiently determine near-optimal solutions for large-scale discrete-choice combinatorial optimization problems. Thus, this article uses one such heuristic, the collective-learning algorithm of I-EPOS~\cite{pournaras2018decentralized,majumdar2023discrete}, for its high performance on plan selection, especially when satisfying hard constraints in a decentralized system (e.g., avoiding monitoring any private region).

The objective of the proposed method is to allow smart cameras, each controlled by a software agent, to autonomously choose their viewing orientations such that the overall coverage is maximized via Eq.(\ref{eq:inefficiency}) while respecting privacy-sensitive regions via Eq.(\ref{eq:constraint_private}). This method achieves coverage optimization for privacy-aware cameras in two complementary ways: plan generation and plan selection.

\cparagraph{Plan generation by individual cameras}
At the beginning, each camera runs the plan generation strategy to independently generate $K$ plans (see Algorithm~\ref{algorithnm1}), with input of location $L_u$, the field of view $F$ and the 2D map information. Depending on the total number of plans $K$ that need to be generated, a camera uniformly discretizes its rotation range $\theta_{uk}$. For example, generating four plans corresponds to evaluating four possible orientations, each spaced $90^{\circ}$ apart (Line 4). Then, given $L_u$, $F$, and $\theta_{uk}$, the camera calculates its coverage ratio $p_{ukn}$ at each cell through Eq.(\ref{eq:field_view}) (Lines 5-6). Next, the camera checks two conditions: (1) whether its entire field of view is outside the map boundaries (Line 7), and (2) whether its field of view covers the private regions (Line 8). Here, we set a threshold value $V$ that permit the peripheral edges of the sectoral field of view to cover the private regions. This can prevent the camera from operating without any plan. If both conditions are not satisfied, the camera generates the plan $p_{uk}$ and adds it to the plan set $\mathcal{K}_u$ (Lines 9-10). 

\cparagraph{Collective learning for plan selection}
After generating their feasible plans, cameras begin the collective learning process to coordinate their choices. The coordination process is as follows: The agents of cameras organize themselves into a tree communication structure (see Fig.~\ref{fig:scenario}), where each interacts only with its children and parent. Through iterative bottom-up and top-down exchanges, cameras share aggregated information about their coverage and gradually improve their plan selections~\cite{pournaras2018decentralized}. In the bottom-up phase, each camera (except leaf cameras) aggregates the plans received from its children and selects its own plan by minimizing soft constraints defined in Eq.(\ref{eq:inefficiency}). The selected plan is then combined with the aggregated plans from its children and forwarded to its parent. This process continues up the tree until the root camera aggregates the information from all cameras and obtains the overall coverage $G$. In the top-down phase, the root camera sends $G$ back through the tree, allowing each camera to evaluate how its current orientation contributes to the match between the overall coverage $G$ and target $T$, and then adjust its plan in the next iteration if a better match can be achieved. Through repeated iterations, cameras refine their orientations to match $G_n$ to $T_n$ at each cell, $\forall n \leq N$, improving coverage in critical regions ($T_n = 1$) while discouraging observation of private regions ($T_n = 0$).

\begin{figure}[!t]
    \centering
    \includegraphics[width=\linewidth]{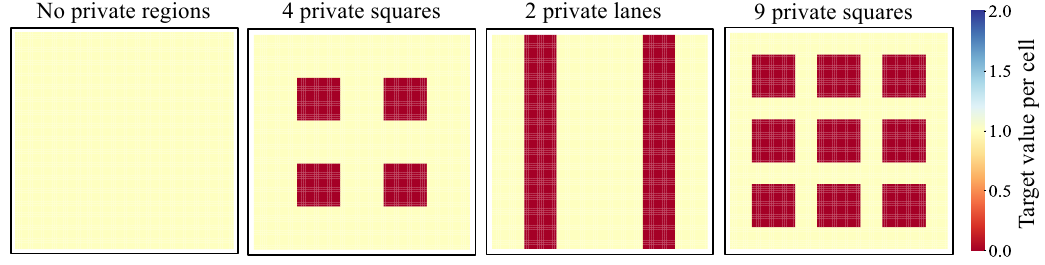}
    \caption{The coverage value per cell required by the targets with and without private regions.}
    \label{fig:targets}
\end{figure}

\section{RESULTS AND DISCUSSIONS}

\subsection{Experimental Settings}
The goal of experimental evaluation is to demonstrate that high coverage efficiency can be achieved using a minimal number of cameras (or minimum budget), while simultaneously ensuring effective coverage exclusion of private regions. To this end, we study a square environment of $1000m \times 1000m$, divided into $N = 10,000$ uniform cells of size $10m \times 10m$. Each smart camera is modeled as a full-frame device with a sensor width of $w_\text{I} =35mm$ and a typical focal length of $f = 31mm$. These parameters are used to calculate the horizontal angle of $\alpha = 45^{\circ}$ and the corresponding field of view $F$ through Eq.(\ref{eq:angle}) and (\ref{eq:field_view}). To study the scalability and adaptability of the proposed method, several key environmental and system parameters are varied, including the number of deployed cameras $U$, the number of orientation plans $K$ available to each camera, the camera locations $L_u$, $\forall u \in \mathcal{U}$, and different types of private regions (see Fig.\ref{fig:targets}).

During the coordinated plan selection via I-EPOS\footnote{I-EPOS is open-source and available at https://github.com/epournaras/EPOS.}, cameras self-organize into a balanced binary tree as a way of structuring their learning interactions~\cite{majumdar2023discrete}. The shared goal of the agents is to minimize the RMSE between the target and overall coverage of cameras, both in unit-length scale. The algorithm repeats $40$ times by changing the random position of the cameras on the tree\footnote{More information about the influence of the tree topology and agents' positioning on the tree is illustrated in earlier work~\cite{pournaras2018decentralized}.}. At each repetition, the cameras perform $40$ bottom-up and top-down learning iterations during which RMSE converges to the minimum optimized value. 

A fair comparison of the proposed method with related work is not straightforward, as there is a very limited number of relevant decentralized algorithms. These algorithms~\cite{altahir2017optimizing,aribilola2022securecam} either do not address privacy concerns, or have higher computational and communication overhead than \emph{I-EPOS} (see Related Work section). For this reason, we use two baseline methods capable of performing multi-camera coverage optimization: \textit{Optimal} and \textit{Greedy Grid Voting} (\textit{GGV}). Here, \emph{Optimal} is a centralized approach that manually adjusts the orientation of all cameras in the map to achieve the maximum coverage (i.e., the minimum loss). \emph{GGV}~\cite{suresh2022efficient} is a two-stage grid voting algorithm for camera coverage optimization. To protect privacy, we improve \emph{GGV} by setting $0$ preference on private regions which requires cameras not to cover them. In addition, the proposed method is divided into two: \emph{I-EPOS}, which only solves soft constraints, and \emph{I-EPOS-HC}, which incorporates \emph{I-EPOS} with the hard constraint satisfaction~\cite{majumdar2023discrete}. 

To evaluate these methods, we use the \textit{coverage inefficiency}, i.e., the soft constraint defined in Eq.(\ref{eq:inefficiency}), as the primary metric. Other metrics such as \textit{privacy violation rate}, \textit{total coverage ratio}, and \textit{interpolated cost} are introduced in the following subsections.

\begin{figure*}[!t]
\centerline{\includegraphics[width=\linewidth]{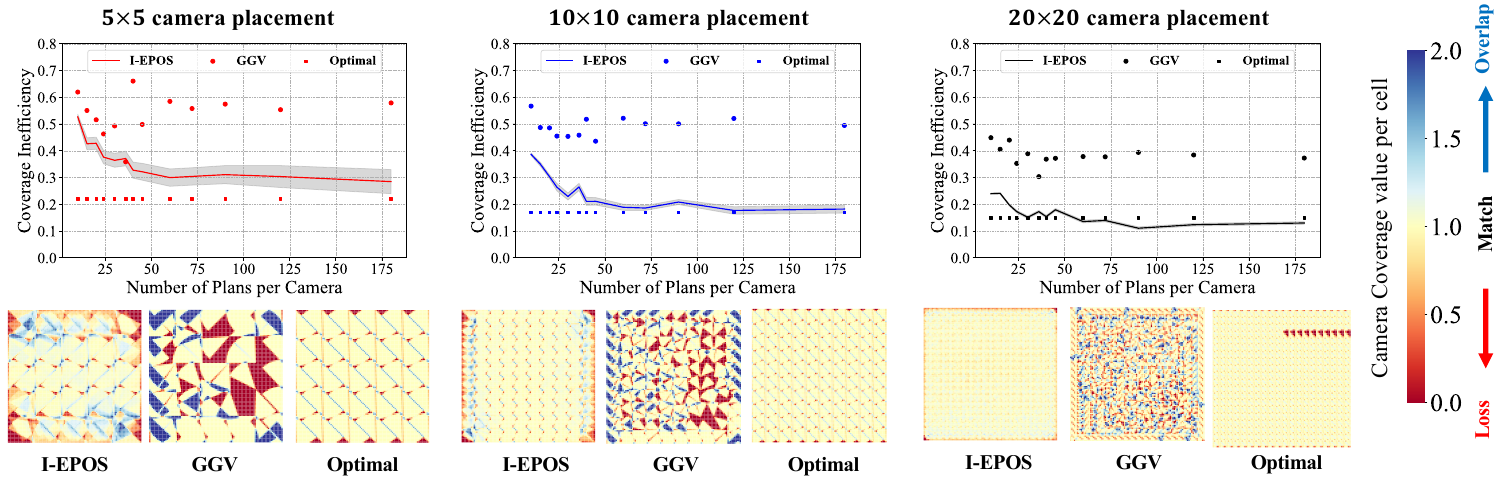}}
\caption{Coverage inefficiency of three approaches with different types of camera placement and number of plans per camera. The shadow represents the standard error of \emph{I-EPOS}. The heatmaps show the coverage performance (match, overlap, or loss) on the 2D map through different methods.}\vspace*{-5pt}
\label{fig:basic_compare}
\end{figure*}

\subsection{Coverage Analysis}
The coverage performance of cameras among different approaches without private regions is studied at first. We increase the number of plans per camera from 10 to 180 and set different types of camera placement, e.g., there are $25$ cameras distributed in $5\times5$. \textbf{Fig.\ref{fig:basic_compare} illustrates that \emph{I-EPOS} has lower coverage inefficiency than \emph{GGV}, particularly with a higher number of cameras and plans.} 

When the number of plans increases, the coverage inefficiency of \emph{I-EPOS} decreases and converges to a minimum value at $90$ plans, which is $45.49\%$ lower than \emph{GGV}. This proves that \emph{I-EPOS} can coordinate cameras with high flexibility to adjust their orientations more effectively compared to \emph{GGV}. As a result, the camera coverage of most cells using \emph{I-EPOS} matches the target, i.e., $G_n = T_n = 1$, effectively avoiding observing the same region (i.e., overlap, $G_n > 1$) and leaving large gaps (i.e., loss, $G_n < 1$).

When the placement of cameras changes from $5\times5$ to $20\times20$, the density of cameras in the map increases from $0.0025$ per cell to $0.04$ per cell. In this setting, \emph{I-EPOS} observes the coverage regions effectively, decreasing coverage inefficiency and error. The results demonstrate the effectiveness of \emph{I-EPOS} that coordinates a high number of cameras: it has $71.93\%$ lower coverage inefficiency than \emph{GGV} at $20\times20$ cameras and $90$ plans. Additionally, \emph{I-EPOS} has only $9.69\%$ higher coverage inefficiency than the expected optimal solution in \emph{Optimal} at $10\times10$ cameras since \emph{I-EPOS} has higher loss areas at the corners (see heatmaps in Fig.~\ref{fig:basic_compare}). Yet, at $20\times20$ cameras, even though \emph{Optimal} can control cameras to cover the whole area, it has higher overlap areas than \emph{I-EPOS}, leading to  a $3.98\%$ higher coverage inefficiency.

\begin{figure*}[!t]
\centerline{\includegraphics[width=\linewidth]{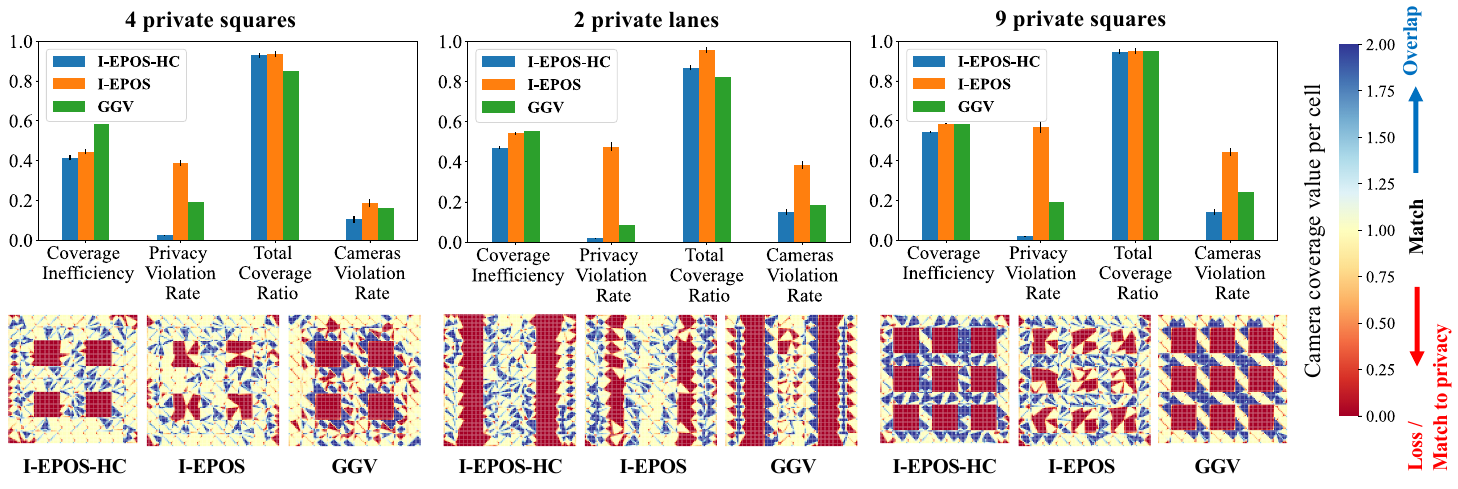}}
\caption{The coverage performance comparison of different approaches using $10 \times 10$ cameras (each with $90$ plans) in the maps with three types of private regions. The metrics include coverage inefficiency, privacy violation rate, and total coverage ratio.}\vspace*{-5pt}
\label{fig:privacy_results}
\end{figure*}

\subsection{Evaluation With Private Regions}
The coverage performance of different methods in the maps with private regions is then studied. As shown in Fig.\ref{fig:targets}, several types of maps with private regions are defined: (i) 4 private square areas ($20\times20$ cells), (ii) 2 private lanes ($15\times100$ cells), and (iii) 9 private square areas ($20\times20$ cells). Moreover, two extra metrics are introduced: (1) \textit{privacy violation rate} indicates the proportion of area covered by all cameras within the privacy regions, (2) \textit{total coverage ratio} denotes the ratio of total area covered by all cameras across all required coverage regions, and (3) \textit{cameras violation rate} denotes the proportion of Cameras that violates privacy.

\textbf{Fig.\ref{fig:privacy_results} illustrates \emph{I-EPOS-HC} deliberately trades a small loss in coverage of non-private areas for a significant reduction in privacy violation, resulting in lower coverage inefficiency compared to \emph{I-EPOS} and \emph{GGV}.} By enforcing privacy via hard constraints, smart cameras using \emph{I-EPOS-HC} automatically adjust their orientations not to observe the private regions, even when such regions occupy a large portion of the map. As a result, it achieves an average reduction in privacy violation rate of $95.33\%$ compared to \emph{I-EPOS} without hard constraints and $85.53\%$ reduction compared to \emph{GGV}. In addition, cameras violation rate of \emph{I-EPOS-HC} is lower than that of \emph{I-EPOS} by $61.08\%$ and \emph{GGV} by $32.48\%$. This strong privacy protection comes at a modest cost. \emph{I-EPOS-HC} discards camera plans that intersect private regions and forgo some coverage opportunities in non-private regions, leading to a $3.38\%$ reduction in total coverage ratio compared to \emph{I-EPOS}. Overall, by prioritizing strict avoidance of private regions as required by the target signal, \emph{I-EPOS-HC} has a $18.42\%$ lower coverage inefficiency than baseline methods, achieving a more efficient balance between coverage quality and privacy preservation.

\begin{figure*}[!t]
    \centering
    \includegraphics[width=\linewidth]{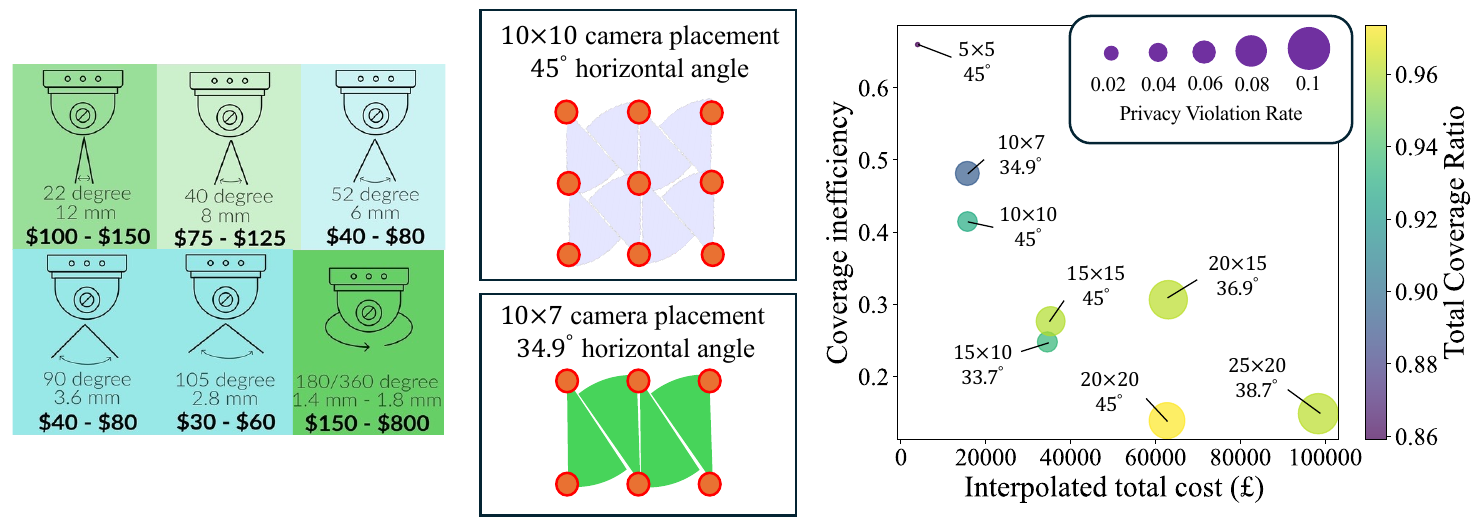}
    \caption{Comparison of interpolated total cost and coverage performance among different types of camera placement. Each type of camera placement requires smart cameras with different horizontal angles and prices according to FIXr.}
    \label{fig:cost}
\end{figure*}

\subsection{Cost Benefit Analysis of Smart Cameras}
In real deployments, camera configurations are constrained not only by coverage quality but also by budget. Different camera placements require different horizontal angles, which in practice correspond to cameras with different unit prices\footnote{Available at: https://www.fixr.com/}. To capture this relationship, we introduce an interpolated cost, which estimates the deployment budget by introducing the price of a standard camera according to the required horizontal angle at each location. The interpolated cost $P_\text{I}$ is formulated as $P_\text{I} = \overline{P} \cdot A^{\text{R}} / A^{\text{S}}$, where $\overline{P}$ is the mean price of a standard camera, $A^{\text{S}}$ indicates the horizontal angle of a standard camera, and $A^{\text{R}}$ denotes the horizontal angle of cameras required by their placement, e.g., $45^\circ$ for $10 \times 10$ cameras and $34.9^\circ$ for $10 \times 7$ cameras which are calculated through \emph{Optimal} (see Fig.~\ref{fig:cost}). In this way, the interpolated cost provides a practical link between abstract coverage performance and the real-world expense of installing smart cameras.

Fig.~\ref{fig:cost} compares the interpolated total cost and coverage inefficiency under different camera placements using \emph{I-EPOS-HC} in the map with 4 private squared areas. Among all evaluated configurations, the $10 \times 10$ placement with $45^\circ$ achieves a Pareto optimal balance between cost and coverage performance. It delivers $0.41$ coverage inefficiency at a relatively modest interpolated total cost, while also maintaining only $0.025$ privacy violation rate. Increasing the deployment to $15 \times 15$ cameras with $45^\circ$ raises both the interpolated total cost and privacy violation rate by around $2.25$ times, but improves the total coverage ratio by only $3.36\%$. This indicates substantially higher expense yields only marginal performance gains. In addition, adopting a rectangular placement $10 \times 7$ with $34.9^\circ$, which requires fewer but higher-priced cameras than $10 \times 10$, results in a similar budget level. However, in $10 \times 7$, the coverage inefficiency increases by $16.09\%$ and the privacy violation rate rises by $50\%$, reflecting less effective coordination under this spatial distribution of cameras. \textbf{Overall, the $\textbf{10} \times \textbf{10}$ camera placement emerges as the most balanced solution, which delivers efficient, privacy-aware coverage without excessive expenditure.}

\section{CONCLUSION AND FUTURE WORK}
In conclusion, this work moves beyond traditional coverage maximization and cryptographic privacy protection, validating how smart cameras can be coordinated in a way that is both technically effective and socially responsible. By combining decentralized collective learning with hard constraint satisfaction, the proposed framework coordinates hundreds of cameras to autonomously select orientations that jointly optimize coverage while proactively avoiding private regions. 

Experimental results show that our approach achieves $18.42\%$ higher coverage efficiency and $85.53\%$ lower privacy violation than baseline methods. They also provide city planners and operators with quantitative guidance on optimal camera deployments by evaluating trade-offs among budget, coverage quality, and privacy risk. Consequently, this work advances privacy-aware smart camera systems that respect citizens’ rights without additional cost and without performance loss. More broadly, it supports the next generation of legitimate urban CCTV systems, where ``smarter'' cameras actively avoid intrusive surveillance, build public trust, and guide beneficial deployments.

In the future work, we aim to enhance the flexibility and scalability of the decentralized coordination framework by expanding the plan space beyond fixed 2D orientations toward richer, fully 3D models using thousands of cameras. Moreover, we envision an open, blockchain-enabled system where users can access video data and report privacy violations, building a decentralized and trustworthy auditing process that enforces privacy protection even against malicious or non-compliant cameras.

\section{ACKNOWLEDGMENTS}
This research is supported by a UKRI Future Leaders Fellowship (MR-/W009560/1): \emph{Digitally Assisted Collective Governance of Smart City Commons–ARTIO}.

\bibliographystyle{unsrt}
\bibliography{reference}

\end{document}